# EFFECT OF PARTICLE SIZE ON THE SHEAR STRENGTH BEHAVIOUR OF SANDS

**Mohammad Nurul Islam[1], Ayesha Siddika[2], Md. Belal Hossain[3], Azrin Rahman[3], Md. Abdullah Asad[3]**
[1]Assistant Professor, [3]Research Scholar, Department of Civil Engineering, Rajshahi University of Engineering & Technology, [2]Instructor, Department of Civil Engineering, Rajshahi Polytechnic Institute Sopura, Rajshahi 6204, Bangladesh

## ABSTRACT

During plastic deformation of granular materials due to loading, the stress-strain and strength characteristics of sand grains are influenced with grain size, their distribution and packing. Also the macroscopic behaviour of granular materials changes with the variation of microscopic behaviour. Particle size is one of the important properties which plays a dominant role on the stress, strain and strength responses of granular materials. Alteration of grain size results in the change of void ratio as well as particle effective contact area revolutionized and the load distribution mechanism of particle to particle contact. To evaluate the effect of particle size, a series of direct shear tests were performed considering *uniform particles* of eight samples (0.075, 0.15, 0.212, 0.300, 0.600, 1.18, 1.72 and 2.76 mm) and *graded particles* of two samples (0.075-1.18 mm and 0.075-2.36 mm). Three types of normal loads (0.05, 0.10 and 0.15 kN) were selected for each test. For uniform particles, particles retained on individual sieve size were considered and in graded particles combination of each uniform particle pondered. A theoretical approach was also proposed to correlate the particle size and macroscopic response. From the experimental results it was observed that for each set of normal load with the increase of particle size, angle of internal friction as well as maximum horizontal shear stress increases for uniform sands and a similar response was noticed in graded sands but the larger the gradation the higher the shear strength. Maximum horizontal shear and angle of internal friction with respect to particle size is also influenced by normal stress. Experimental results have good agreement with the theoretical approach.

## 1     INTRODUCTION

Particle size plays an important role on the strength behaviour of granular materials. The size of the particles in the granular mass alters the fabric and is responsible forthe variation of strength behaviour. When granular materials are subjected to loading, particle size and their associated voids play an important role in the dissipation of energy. The load transformation mechanism of the granular mass depends on the individual soil grains as load transfer particle to particle and the macroscopic response of granular mass is the resultant of the individual response of the particles.

To evaluate the influence of particle size, numerous laboratory as well as numerical tests have been performed in the last few decades. The diversity of the conclusions of conducted research revealed an ambiguity which lead to further investigations. Moreover, theoretical relationships between the particle size and stress-strain and strength are still elusive. In laboratory experiment, different types of testing apparatus (direct shear test, triaxial test etc.) and in numerical analysis 2D-DEM and 3D-DEM (Discrete Element Method) were adopted. Frederick (1961) pointed out that macroscopic response of granular materials changed with the alteration of particle size. Kolbuszewski and Frederick (1963) reported that macroscopic behaviour of the granular mass is the summation of the individual response of the particles and the deformation of the mass before crushing depends on the individual soil grain size. In order to investigate the effect of particle size, Kolbuszewski and Frederick (1963) used ballotini (glass beads) and concluded that maximum porosity decreases and compressibility increases with the increase of particle size. Particle size effects dilatancy component during the triaxial test. Zolkov and Wiseman (1965) conducted triaxial tests to study the effect of particle size and observed that the angle of shearing resistance increases as particle size increases. Kolbuszewski and Frederick (1963) reported similar conclusions to Zolkov and Wiseman (1965) but Kirkpatric (1965) found a different result. According to Kirpatric (1965) the angle of shearing resistance decreases with the increase of particle size, which agreed with the findings reported in Marschi *et al.,* 1972. To determine the effect of particle size Chattopadhya (1981) performed direct shear tests using Mogra sand and described that the fabric of the granular mass is a function of the particle sizes and their distribution and the maximum void ratio increase with the decrease of particle size while Kirpatric (1965) reported maximum and minimum void ratios remain constant with the particle size. From the numerical observations (2D-DEM) Sitharam (2000) suggested that due to change of the particle size and gradation, stress-strain behaviour of granular mass changes. A change of particle gradation maintaining the minimum particle size as constant, results in the decrease in the angle of internal friction and also increase in volumetric strain. Martinez (2003) using conventional triaxial test reported that particle size affect the stress-strain and volumetric strain behaviour of the granular materials. Katzenbach and Schmitt (2004) using 3D-DEM (PFC) conducted triaxial simulation to evaluate particle size effect and found that particle size distribution strongly alters the stress-strain behavior. Herbold *et al.* (2008) reported that the dynamic mechanical properties alter with the increase of particle size. Gupta (2009) investigated particle size (25, 50 and 80 mm) for Ranjit Sagar Rockfill Material (RSRM) and Purulia Rickfill Material (PRM) and found that for RSRM angle of internal friction increase with the increase of particle size for RSRM but for PRM angle of internal friction decrease with the increase of particle size for PRM. Islam (2009) using 3D-DEM (YADE) observed that particle size alters the strength characteristics.

Table 1: Summary of the Review of the Effect of Particle Size

| Year | Researcher | Sample | Test | Findings |
|---|---|---|---|---|
| 1961 | Frederick | Sand | TT | Particle size is responsible for change in the macroscopic response. |
| 1963 | Kolbuszewski and Frederick | Glass beads | TT | Angle of shearing resistance increase with the increment of particle. Size |
| 1965 | Zolkov and Wiseman | Sand | TT | Shear strength increase with respect to particle size. |
| 1965 | Kirkpatric | Sand | TT | Shearing strength decrease with the increase of particle size. |
| 1972 | Marschi | Sand | TT | Angle of internal friction decrease with increase of particle size. |
| 1981 | Chattopadhya | Sand | DST | Maximum void ratio increases with the decrease of particle size. |
| 2000 | Sitharam | 2D-DEM | BT | Particle size and gradation alters the stress-strain behavior. |
| 2003 | Martinez | Glass beads | TT | Particle size affects the stress-strain and volumetric response of granular materials. |
| 2004 | Katzenbach and Schmitt | 3D-DEM | TT | Particle size distribution strongly alters the stress strain behaviour. |
| 2009 | Gupta | Rock | TT | Angle of internal friction increase with increase of particle size for Ranjit Sagar Material but for Purulia Rockfill Materials opposite value observed. |
| 2009 | Islam | 3D-DEM | TTT | Stress, strain and strength behaviour significantly change when particle size change even though void ratio remain constant. |

*Note:* TT = Triaxial Test, TTT = True Triaxial Test, BT = Bi-axial Test, DST = Direct Shear Test

## 2    SAMPLE PREPARATION

To investigate the effect of particle size, two types of samples were selected: *uniform particle* (particles of about same size) and *graded particle* (particles of wide range of size). For uniform particle, eight (8) samples (Figure 1: *a* to *h*) were considered, which can be divided under three headings according to Australian Standards AS1726-1993: *Fine Sand* (0.075 mm, 0.150 mm), *Medium Sand* (0.212 mm, 0.300 mm, 0.600 mm) and *Coarse Sand* (1.18 mm, 1.72 mm and 2.76 mm). For graded particles two types of gradation were selected: 0.075 to 1.18 mm and 0.075 to 2.36 mm. The grain size distribution of the graded samples is shown in Figure 2 along with calculated uniformity coefficient ($C_u$) and curvature coefficient ($C_g$).

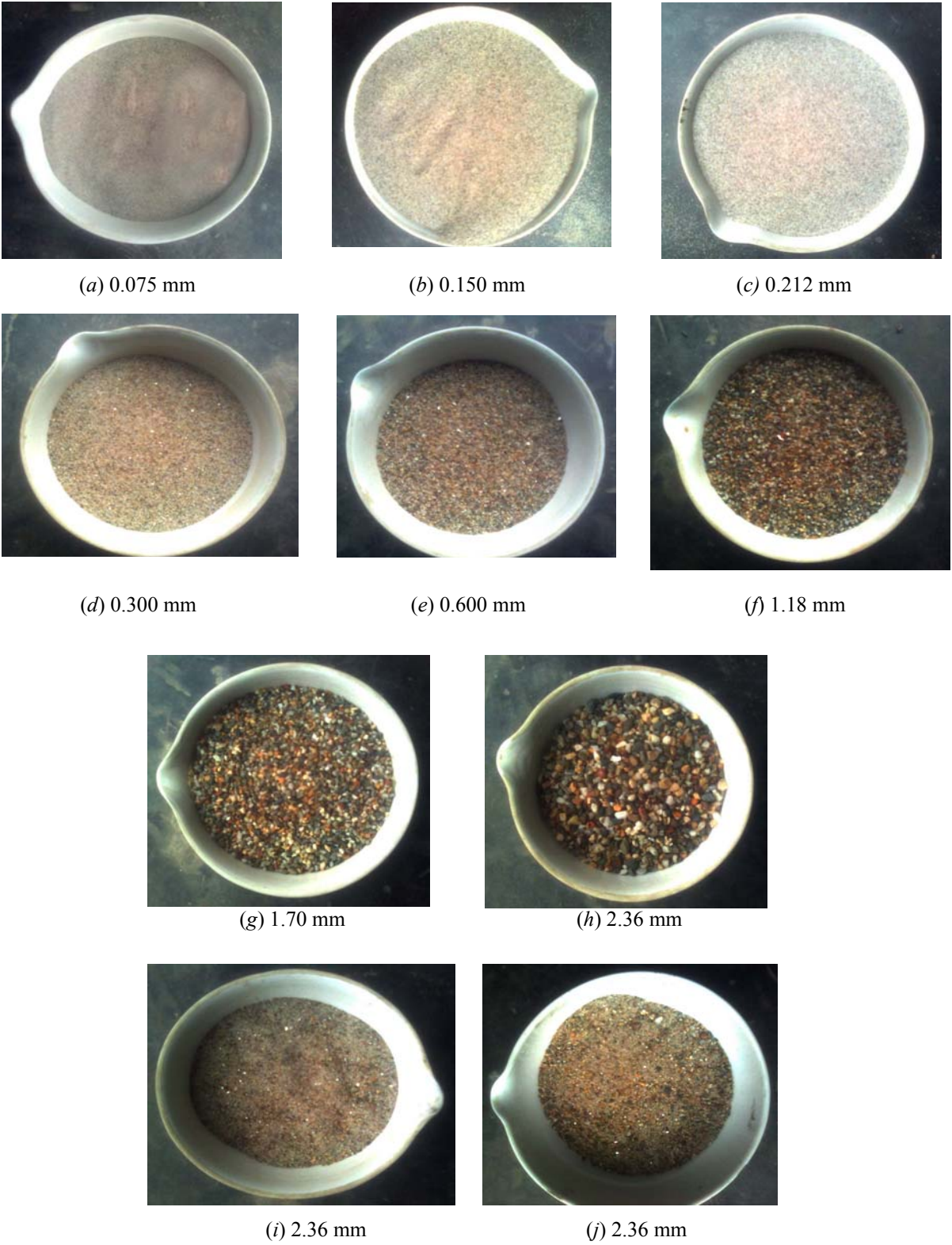

Figure 1: Samples used in Test: Uniform Particles (*a* to *h*) and Graded Particles (*i* to *j*)

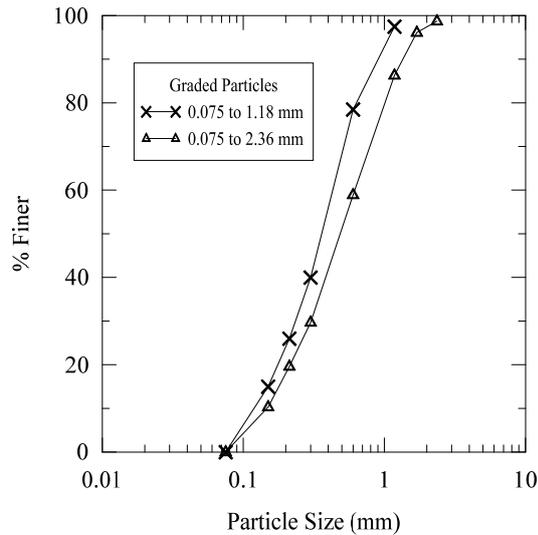

| Items | 0.075 to 1.18 mm | 0.075 to 2.36 mm |
|---|---|---|
| $D_{10}$ | 0.12 | 0.14 |
| $D_{30}$ | 0.24 | 0.31 |
| $D_{50}$ | 0.36 | 0.43 |
| $D_{60}$ | 0.42 | 0.62 |
| $C_u$ | 3.50 | 4.43 |
| $C_g$ | 1.14 | 1.10 |

*Note*:

$C_u$ = Uniformity Coefficient = $\dfrac{D_{60}}{D_{10}}$

$C_g$ = Curvature coefficient = $\dfrac{D_{30}^2}{D_{60} \times D_{10}}$

$D_{10}, D_{30}, D_{50}, D_{60}$ = Grain size (mm) corresponding to 10, 30, 50 and 60% passing

Figure 2: Gradation of Particle Size.

Table 2: Properties of the Sand

| Particle size(mm) | Void ratio | Water content (%) | Unit weight (gm/cm³) | Angle of Internal Friction | Maximum Shear Strength (kPa) | | |
|---|---|---|---|---|---|---|---|
| | | | | | 0.05 kN | 0.10 kN | 0.15 kN |
| 0.075 | 0.95 | 0.26 | 1.29 | $35.54^0$ | 11.06 | 24.33 | 32.56 |
| 0.150 | 0.90 | 0.18 | 1.33 | $36.87^0$ | 11.50 | 24.99 | 33.40 |
| 0.212 | 0.87 | 0.21 | 1.34 | $37.57^0$ | 11.94 | 26.54 | 35.39 |
| 0.300 | 0.83 | 0.22 | 1.36 | $38.30^0$ | 12.38 | 26.76 | 35.39 |
| 0.600 | 0.74 | 0.21 | 1.37 | $38.93^0$ | 13.27 | 27.42 | 36.49 |
| 1.18 | 0.56 | 0.16 | 1.48 | $40.82^0$ | 14.37 | 28.53 | 37.16 |
| 1.70 | 0.51 | 0.22 | 1.52 | $41.19^0$ | 15.04 | 29.64 | 38.04 |
| 2.36 | 0.38 | 0.31 | 1.60 | $42.27^0$ | 15.48 | 29.86 | 38.48 |
| 0.075-1.18 | 0.49 | 0.27 | 1.46 | $41.18^0$ | 14.60 | 28.98 | 38.49 |
| 0.075-2.36 | 0.54 | 0.32 | 1.51 | $41.83^0$ | 15.04 | 29.64 | 39.15 |

All samples (uniform as well as graded) were natural sand (NS) rather than manufactured or crushed sand (CS). According to (Villalobos *et al.*, 2005; Cho *et al.*, 2006) the NS and CS differ in particle shape and surface characteristics. The NS are rounded whereas CS is angular, the void ratio of NS compared to CS is lower where fineness of CS is higher compared to NS. The sand particles are locally available from the Padma River Sand in Rajshahi. In nature, it is difficult or even impossible to determine the individual particle sizes or mono size particles. But a numerical model like Discrete Element Method (DEM) is capable of deal mono size particles. The author investigated the mono sized particles (Islam, 2009) to evaluate the effect of particle size considering the b-value and found particle size also plays an important role. In sample preparation, particle size was selected based on the amount of soil particle passing through a specific sieve opening but retained on a sieve of smaller sized openings. Soil grains retained on any sieve in this way was the combination of specific particles size range (not mono size particle), all of which were smaller than the openings of the sieve through which the material passed but larger than the openings of the sieve on which the soil was retained. The shape of the particles was almost rounded. The properties of the sand grains which were used to investigate the effect of particle size are tabulated in Table 2.

Specimens were prepared using the compaction method. For sample preparation ASTM D3080 was followed. A mould with a diameter of 5.08 cm was used to prepare the specimen. Each sample was poured into the mould in three layers and tamped to compact the materials. Before starting the tests, samples were submerged in water to avoid the fine content as fine content has influence on the shear strength behaviour of granular materials (Rahman, 2009). Then the samples were air dried for 24 hours prior to oven drying for 12 hours.

## 3 TEST PROCEDURE

The direct shear test was conducted by deforming the specimen until failure. At failure maximum shear stress was obtained. The test was performed by applying a predetermined normal stress (*0.05 kN, 0.10 kN and 0.15 kN*). In the direct shear test there is a device which holds the specimen securely between two porous inserts. Care was maintained so that torque was not developed or applied to the specimen. The shear device is capable of applying a shear force to the sample along a specified shear plane, which is parallel to the faces of the specimen. There are shear boxes of circular section (5.08 cm diameter) in a direct shear test. The box is divided vertically by a horizontal plane into two halves. Porous inserts are placed in top and bottom of a sample. It transfer horizontal shear stress from the insert to the top and bottom boundaries of the sample. There are two types of loading device: loading device for applying and measuring normal forces and loading device for shearing the specimen. For the measurement of shear force a proving ring or load cell was used. During the shear the strain rate was uniformly applied throughout the tests.

## 4 THEORETICAL APPROACH

When sand grains in granular mass touch each other at finite contacting area (dry condition), the external stresses are carried by the particulate mass. Due to excitation of external force, particles move. The shearing deformation of granular mass under Triaxial compression originated along the sliding plane (Figure 3), where $\left(\dfrac{\tau}{\sigma}\right)_{max}$ is developed (Murayama, 1964).

According to Murayama (1964) granular soil accumulates along the sliding plane due to application of load. If defined granular sand among the accumulated one is considered and the contacting point is marked by X and Y (Figure 3) on the top of the sliding plane the defined grains are excited by the external forces and the resultant force is expressed by $(p_r)_i$. The resultant force $(p_r)_i$ composed of the normal force $(p_n)_i$ (caused due to normal stress, $\sigma$) and shear force $(p_s)_i$ (caused due to shearing stress $\tau$). $(\gamma)_i$ is the angle between resultant force $(p_r)_i$ and normal to the sliding plane. With the increase of normal stress $\tau$, the angle $(\gamma)_i$ increase and the contact pressure at Y becomes smaller and sand grains begins to slide only touching at point X. The angle of individual sliding surface against sliding plane is $(\alpha)_i$ and the sliding condition of defined grain is $[(\gamma)_i - \mu > (\alpha)_i]$, $\mu$ is frictional angle. $[(\gamma)_i - \mu]$ vary at every grain and subjected to specified frequency distribution.

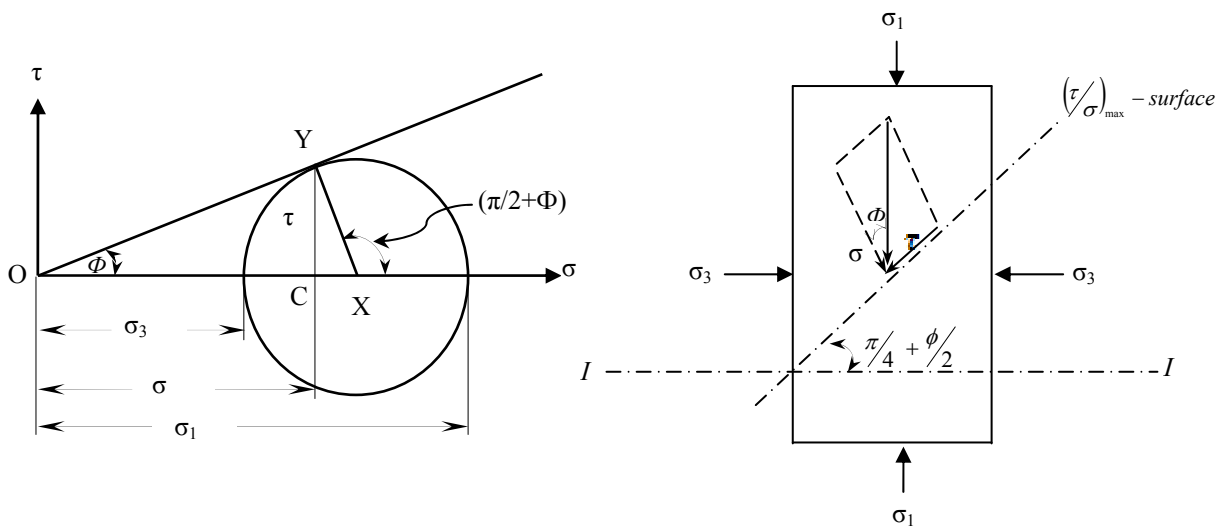

Figure 3: Sliding plane where $(\tau/\sigma)$ is the maximum under triaxial compression

Let $\gamma$ and $\alpha$ be the arithmetic means of $(\gamma)_i$ and $(\alpha)_i$ respectively and N is the total number of granular sand (in a finite volume) then $\gamma$ and $\alpha$ can be expressed by Equation 1

$$\gamma = \sum_i^N \frac{(\gamma)_i}{N} \quad , \quad \alpha = \sum_i^N \frac{(\alpha)_i}{N} \tag{1}$$

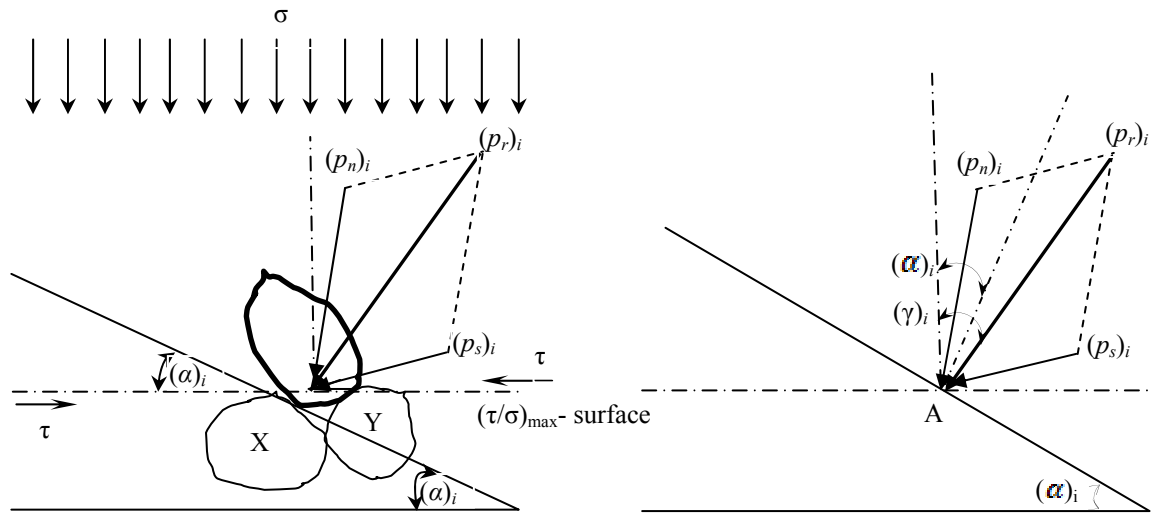

Figure 4: Forces acting on sand grain moving along the surface of the adjacent grain. (After Murayama, 1964).

In nature granular sands size and shape are random and the direction and magnitude of $(p_n)_i$ and $(p_s)_i$ are random. The average forces $(p_n)$ and $(p_s)$ can be expressed by Equations 2 and 3 respectively (Figure 4).

$$p_n = \sum_i^N (p_n)_i \cdot \cos\left[(p_n)_i \wedge \sigma\right] / N = \sigma / N \tag{2}$$

$$p_s = \sum_i^N (p_s)_i \cdot \cos\left[(p_s)_i \wedge \tau\right] / N = \tau / N \tag{3}$$

$$\text{As } \gamma = \tan^{-1}\left(\frac{p_s}{p_n}\right), \text{ and}$$

$$\gamma = \tan^{-1}\left(\frac{p_s}{p_n}\right) = \tan^{-1}\left(\frac{\tau}{\sigma}\right)_{max} = \phi \tag{4}$$

From the Equations 1, 2, 3 and 4 Equations 5 becomes

$$\left.\begin{array}{l} \alpha = f(\alpha_i, N) \\ \gamma = f(\gamma_i, N) \\ p_n = f(\sigma, N) \\ p_s = f(\tau, N) \end{array}\right\} \tag{5}$$

Let the volume of solid is $V_s$, volume of voids is $V_v$ and total number of particle per unit area to be $N$. Then total volume, $V = V_v + V_s$ Where $V_s$ = Volume of solid particles = $N\frac{4}{3}\pi r^3$ The number of particles becomes,

$$N = \frac{V_s}{\frac{4}{3}\pi r^3} \tag{6}$$

where $r = \frac{r_{max} + r_{min}}{2}$ = Average radius, but in practical, $r$ is different.

Again, Total Volume $, V = V_S(1+e)$

Where $V$ = Total volume and $V_s$ = Volume of Solids, $e$ = Void ratio

$$\text{Then } V_s = \frac{V}{1+e} \tag{7}$$

The volume per particle will be (for unit section),

$$V = 1 \times d^3 = (2r)^3 = 8r^3$$

So total volume will be
$$V = 8Nr^3 \tag{8}$$

From Equation 7 and 8

$$V_s = \frac{8Nr^3}{1+e}$$

$$\therefore N = \frac{V_s(1+e)}{8r^3} \tag{9}$$

From Equation 5 it was observed that α, γ, $p_n$, $p_s$ depends on particle number, N and particle number is the function of particle size, *r*. On the other hand, void ratio *e* is the function of particle radius *r*. So particle size (microscopic properties ) plays an important role on the macroscopic response.

## 5      EFFECT OF PARTICLE SIZE

To investigate the effect of particle size, three sets of loading (0.05, 0.10 and 0.15 kN) were selected for each of eight particle sizes. For each load set it was observed that with the increase of particle size, the shear strength properties increased as shown in Figure 5, which shows that particle size influences shear strength behaviour. Selig and Roner (1981) using triaxial tests reported that for varying void ratio with the increase of the particle size, shear strength increases and void ratio decreases with an increase of particle size for constant volume, but the shear strength is independent for common void ratio. Islam (2009) using 3D-Discrete Element Method (Constant Mean Pressure Test and Constant $\sigma_3$) observed that if void ratio remains constant and particle size changes then shear strength also varies, which is similar to the present experiment (Direct Shear Test). The authors also investigated using the Digital Direct Shear test and observed similar results (Islam *et al*., 2011).

From Figure 5(a), 5(b) and 5(c), it is observed that for each loading (0.05, 0.10 and 0.15kN) with the increment of particle size gradually increases the shear stress. The comparison of maximum shear stress for individual grain size with respect to variation of loading is shown in Figure 5(d) and for three types of loading similar qualitative response is observed. In Figure 5(e) for individual particle size the angle of internal friction (Equation 10) is shown and shows an increase in friction angle with increase in particle size, which is similar to Chattopadhya (1981).

To investigate the effect of gradation, two types of gradation (smaller gradation = 0.075 to 1.18 mm and wider gradation = 0.075 to 2.36 mm were considered and noticed that with the increase of gradation, shear strength increase (Figure 6: *a*, *b* and *c*) and for each gradation normal load effect are shown in Figure 7: *a* and *b* and found that with the increase of normal load shear strength increase like uniform particles.

The authors have also investigated the effect of particle shape using natural grains and strain rate and relative density  considering the same particle size (uniform and graded particles) using the strain rate controlled Digital Direct Shear Test (Islam *et al.,* 2011).

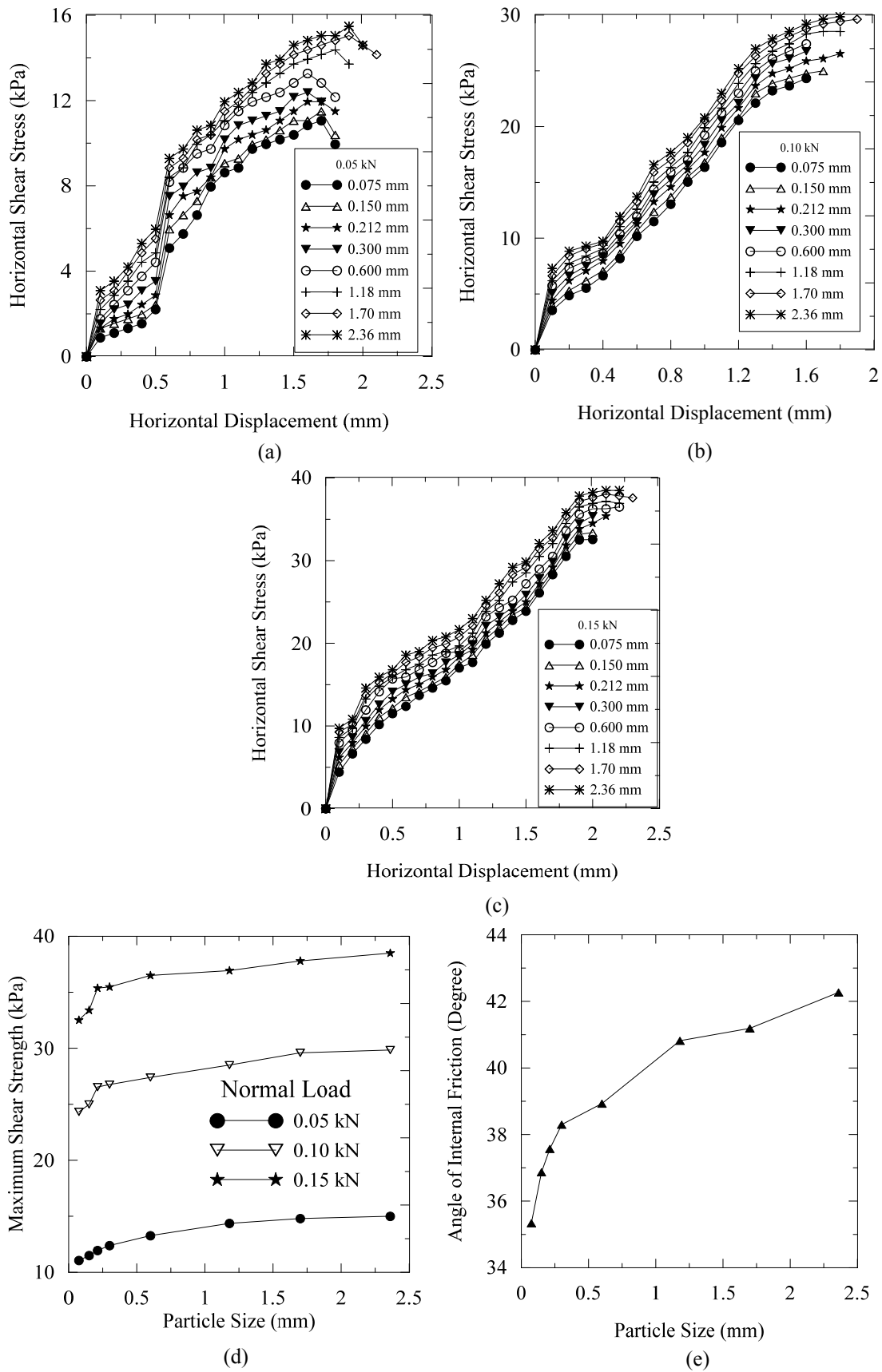

Figure 5: Uniformly graded samples: (a) effect of particle size: 0.05 kN, (b) 0.10 kN and (c) 0.15 kN, (d) effect of Normal load on maximum shear stress and (e) effect of angle of internal friction.

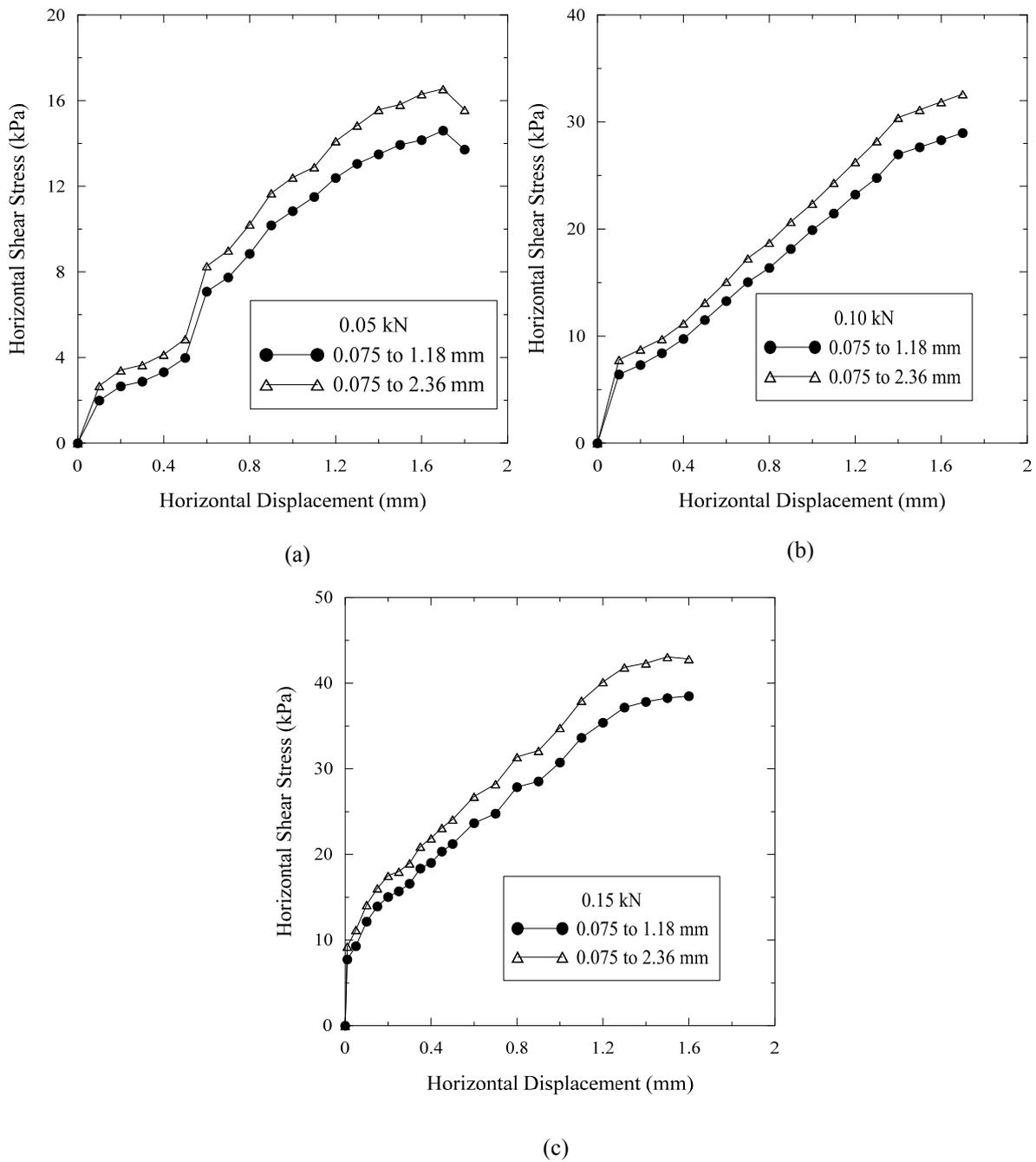

Figure 6: Effect of Gradation in well graded samples (*a*) 0.05 kN, (*b*) 0.10 kN and (*c*) 0.15 kN

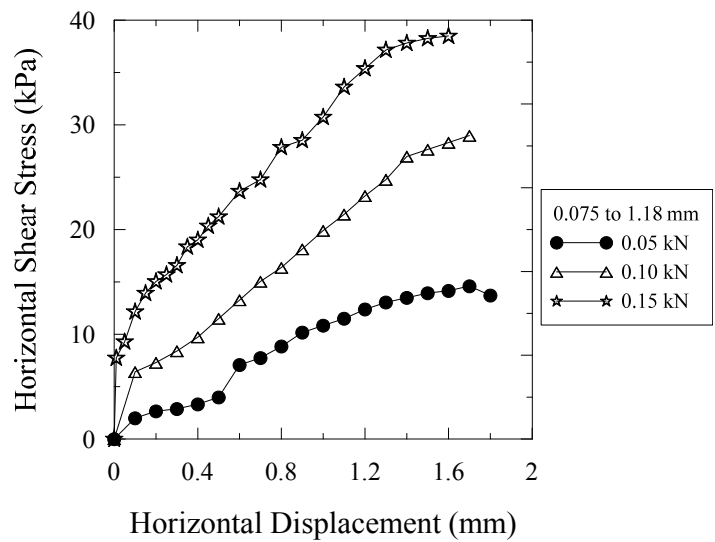

(a)

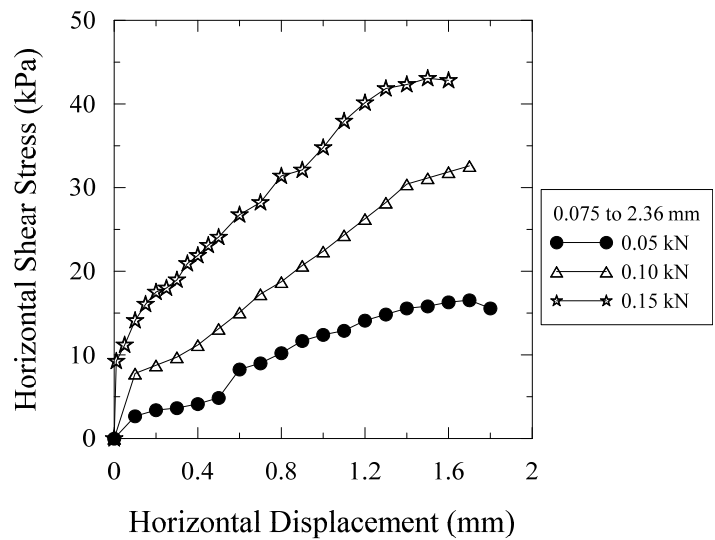

(b)

Figure 7: Effect of Normal Load in well graded samples (*a*) 0.075 to 1.18 mm and (*b*) 0.075 to 2.36 mm

## 6    CONCLUSIONS

Due to change of particle sizes, associated void space among the grain mass alters and changes the contact surface. On the other hand, the variation of particle size influences the rolling resistance and sliding resistance and is also responsible for the deviation of the shear strength behaviour of the granular media. As a result, when the granular body is subjected to load, the stress, strain and strength responses also differ with particle size.

In the present study, the effect of particle size was investigated and the outcomes from the experimental results are as follows:

- With an increase of particle size, the maximum shear strength as well as angle of internal friction increases and the normal load also plays important role.
- With an increase of gradation smaller to wider, maximum shear strength and angle of internal friction increase. The response of maximum shear strength is higher in graded particles as compared to uniform particles.
- In a theoretical approach the relationship between particle size and force has been developed. From the mathematical relation it is observed that normal stress, shear stress and the resultant force alters due to change of particle size, which also satisfies the experimental results.